\documentclass[aps,prc,reprint,nofootinbib,floatfix]{revtex4-2}
\usepackage{float}

\usepackage{graphicx}
\usepackage{dcolumn}
\usepackage{bm}
\usepackage{amsmath}

\usepackage{orcidlink} 
\usepackage{booktabs}
\usepackage{braket}
\usepackage{float}
\usepackage{multirow}

\begin{document}


\title{Modelling fission with microscopic input: excitation-energy-dependent fission paths for neutron-induced reactions}

\author{A. Sánchez-Fernández$^{1,2}$}
\author{W. Ryssens$^{1,2}$}
\author{S. Goriely$^{1,2}$}

\affiliation{$^1$Institut d'Astronomie et d'Astrophysique, Université Libre de Bruxelles, Brussels, Belgium} 
\affiliation{$^2$Brussels Laboratory of the Universe - BLU-ULB, Brussels, Belgium} 

\date{\today}

\begin{abstract}
To calculate transmission coefficients, current Hauser-Feschbach reaction codes assume a one-dimensional fission path that is both independent of excitation energy and universal across different fission channels. In contrast, microscopic fission studies typically explore multiple collective variables. Predicting observable fission quantities such as cross sections based on microscopic input thus requires a way to reduce the dimensionality of the latter. For spontaneous fission, the least action path (LAP) -- which  minimizes the semiclassical action and maximizes the transmission coefficient -- is the solution. We demonstrate that the LAP does not generalize to other fission modes such as neutron-induced or $\beta$-delayed fission: at finite excitation energy, the LAP generally does not maximize the transmission. A microscopic PES can have multiple coexisting stationary-action paths. We show that these (i) can switch their action ordering as a function of excitation energy and (ii) yield different neutron-induced fission cross sections predictions, yet match known spontaneous fission lifetimes equally well.  For the BSkG3 Skyrme-type energy density functional model, the transition from the axially symmetric LAP at zero excitation energy to a path exploiting the triaxial degree of freedom occurs at an excitation energy that is smaller than the neutron separation energy for essentially all actinides and the majority of unknown exotic nuclei.                                                                     .
\end{abstract}

\maketitle

\textit{Introduction} -- Nuclear fission is a complex phenomenon in which a nucleus splits into two or more fragments. Since its discovery, fission has become relevant in many fields, from energy production and nuclear technology~\cite{Nupecc22,aliberti2006} to nucleosynthesis processes~\cite{Goriely15}. Nevertheless, decades of experimental and theoretical efforts have yet to produce  fully predictive microscopic description of fission~\cite{Bender20}.

Fission is most often modelled as a tunneling process: depending on its energy, the nucleus moves over or through an energy barrier that separates compact shapes from scission configurations in a low-dimensional collective space. To connect this picture to the fission cross sections relevant to most applications, the transmission coefficients associated with different fission channels are fed into Hauser-Feschbach simulations. Reaction codes such as TALYS \cite{Koning23}, COH-3 \cite{kawano2021}, CCONE \cite{iwamoto2016}, and EMPIRE \cite{herman2007} base themselves on transition state theory~\cite{eyring1935,bohr1939a,hill1953} and require three ingredients to determine transmission coefficients~\footnote{See Refs.~\cite{bertsch2022,weidenmuller2022,weidenmuller2024a,weidenmuller2024b} for an intriguing microscopic reimagining of transition state theory that has yet to be deployed in practice. }. The first two are the \emph{potential energy} $V(s)$ and the \emph{collective inertia} $\mu(s)$ of the nucleus as a function of a single collective coordinate $s$. Together, $V(s)$ and $\mu(s)$ specify a one-dimensional \emph{fission path}. Third is the \emph{saddle-point nuclear level density} (NLD) $\rho_{s}(E,J,\pi)$, which counts the number of states as a function of energy $E$ with given quantum numbers $J, \pi$ that are accessible for a nucleus near the top of the path. From these three quantities, reaction codes rely on the semiclassical Jeffreys-Wentzel-Kramers-Brillouin (JWKB) approximation to calculate a transmission coefficient $T(E_{\rm exc})$ which depends on the excitation energy $E_{\rm exc}$
of the fissioning nucleus. This transmission coefficient is then used to model (at least) three different fission processes with corresponding energy regimes: (i) spontaneous fission (SF) at  $E_{\rm exc} = 0$, (ii) neutron-induced fission (nF) at $E_{\rm exc} = S_n + E_{\rm n}$, i.e. the excitation energy of the compound nucleus is equal to its neutron separation energy plus the energy of the incoming neutron and (iii) $\beta$-delayed fission ($\beta$DF), where the daughter nucleus has an excitation energy in the $Q_{\beta}$ window, $0 \leq E_{\rm exc} \leq Q_{\beta}$.

Most applications today rely on highly parameterized formulas for these ingredients. Even if such an empirical approach is effective at fitting experimental data~\cite{plompen2020,shibata2011,capote2018,Capote2009}, a microscopic derivation of the structure inputs would (a) be more appropriate for extrapolations to neutron-rich nuclei or other extreme regimes and (b) advance our understanding by linking fission to other nuclear observables. As ab initio approaches struggle to tackle heavy nuclei~\cite{hergert2020} and/or large deformation~\cite{scalesi2026}, mean-field calculations based on nuclear energy density functionals (EDF) are arguably the most microscopic tool capable of predicting the fission properties of thousands of heavy nuclei~\cite{Schunck16}. Today, EDF calculations with a single constraint on the nuclear quadrupole deformation $\beta_{20}$ are routine; the resulting fission paths ($V(\beta_{20}), \mu(\beta_{20})$) are often compared to fission quantities such as empirical barrier heights~\cite{ling2020,taninah2020,Kortelainen12,Rodriguez14,ryssens2019a,Ryssens23} and the corresponding SF half-lives are regularly compared to experiment~\cite{erler2012,Giuliani13,Sadhukhan14,Lemaitre2018,Sanchez2026,Sanchez2026b}. If one accepts a renormalization on existing data, it has been shown \cite{Goriely2009b,Goriely11} that feeding reaction codes with EDF predictions results in good agreement with experiment; this state of affairs is bound to improve if we leverage further progress in the EDF-based modelling of fission~\cite{schunck2016,Ryssens23,Sanchez2026}.

  Unfortunately, there is a mismatch between reaction and structure theory: reaction codes operate in terms of a \emph{one-dimensional} path while structure studies today produce potential energy surfaces (PES) and the associated inertia tensor as a function of multiple collective variables. This is neither new nor specific to EDFs: both microscopic-macroscopic and EDF-based calculations have repeatedly demonstrated the impact of other collective variables \cite{berger1984,moller2009,younes2009,kowal2010,abusara2010,schunck2014,chai2018,chi2023,Ryssens23,ling2020,bernard2019,rodriguez-guzman2022,zdeb2025}. For SF, there is a widespread solution to this problem: one searches for the fission path that maximizes the transmission and the resulting \emph{least action path} (LAP) effectively reduces the tunneling problem to a single dimension.

  In this Letter, we show that the mismatch between reaction codes and their input remains problematic at finite excitation energy,  i.e.~for all fission processes except SF. We then show that even a naive accounting for finite excitation energy results in a striking energy dependence of the LAP\footnote{To the best of our knowledge, this was first observed in practical calculations in Ref.~\cite{Lemaitre2018}.}: two coexisting paths whose action-ordering switches as a function of excitation energy and that result in dramatically different nF cross sections.


\emph{Reaction modelling} -- The JWKB approximation for the transmission probability $P_j$ through a \emph{single-humped} barrier -- specified by a potential energy $V(s)$ and an inertia $\mu(s)$ as a function of a collective coordinate $s \in [s_{\rm min}, s_{\rm max}]$ -- for a wave packet with excitation energy $E_{\rm exc}$ below the energy of the saddle point, i.e.~$E_{\rm exc} \leq \text{max}_s \left[V(s)\right] \equiv B_j$, is given by\footnote{Beware the identical sign errors in the expressions for $P(E_{\rm exc})$ in Refs.~\cite{Goriely2009b,herman2007}.}
\begin{equation}
  P_j(E_{\rm exc})=\frac{1}{1+\exp(2K_j)},
  \label{eq:transmission_probability}
\end{equation}
where the semiclassical action $K_j$ is given by
\begin{equation} \label{eq:TALYS_integral}
  K_j= \frac{1}{\hbar}\int_{a_j}^{b_j} ds \, \left[ 2\mu(s) (V(s) - E_{\rm exc}) \right]^{1/2}.
\end{equation}
The limits of integration $a_j$ and $b_j$ are the classical turning points, i.e., the points on the path surrounding the hump where $V(a_j) = V(b_j) = E_{\rm exc}$.
The JWKB formalism can be extended to excitation energies above the barrier, but its accuracy decreases with increasing $E_{\rm exc}$~\cite{Goriely2009b}. In fact, TALYS reverts to the analytical Hill-Wheeler result for a parabolic barrier when $E_{\rm exc} > B_j$~\cite{Goriely2009b}:
\begin{equation}
  P^{\rm HW}_j = \Big( 1 + \exp \left[ 2 \pi \left(  B_j - E \right)/ \hbar \omega\right]  \Big)^{-1}.
  \label{eq:P_hillwheeler} 
\end{equation}
The parameter $\hbar\omega = \sqrt{2 a \mu_s^{-1}}$ depends on the curvature $a$ of $V(s)$ and the inertia $\mu_{\rm s}$ at the saddle point.

However, the \emph{internal} nuclear degrees of freedom (i.e., all except for $s$) must be accounted for. Transition state theory provides a ``lumped transmission coefficient'' $T_j(E_{\rm exc}, J, \pi)$:
\begin{align}
  T_j(E_{\rm exc},J,\pi) =& \sum_{k} P_j(E_{\rm exc} - E_k(J,\pi))  \nonumber \\
                        &+ \int_{E_d} ^{E_{\rm exc}} d\epsilon \, P(E_{\rm exc} - \epsilon) \rho_s(\epsilon, J, \pi) \, .
\label{eq:transmission_lumped}
\end{align}
In this expression, the discrete transition states at the saddle point with angular momentum $J$ and parity $\pi$ are characterized by excitation energies $E_{k}(J,\pi)$, measured with respect to $B_j$. Above the last discrete state with $E_d = \text{max}_k \left[ E_k(J,\pi) \right]$, a continuum of discrete states with energy $\epsilon$ (again, relative to $B_j$)  and saddle-point NLD $\rho_s(\epsilon, J,\pi)$ contributes up to an energy $E_{\rm exc}$ - see Fig. 1 in Ref.~\cite{hilaire2003} for a depiction.

For actinides, fission paths feature at least two humps and one isomeric well. In that case, additional modelling is needed to calculate the \emph{total} transmission coefficient $T$ for tunneling through both barriers. The most advanced approach to treat the coupling between humps today is the optical model for fission~\cite{herman2007,sin2008}, but this approach has not yet been universally adopted in reaction codes. Discarding the influence of any isomeric wells with the full-damping approximation is much simpler and results in the following expression for a fission path with $N_{\rm h}$ humps:
\begin{align}
  T^{-1}(E_{\rm exc}) = \sum_{j=1}^{N_{\rm h}} T_j^{-1}(E_{\rm exc}) \, .
\label{eq:full_damping}
\end{align}
Equations~\eqref{eq:transmission_probability} - \eqref{eq:full_damping} are typically discussed under the tacit assumption that $E_{\rm exc}$ is sizeable, typically at least larger than the excitation energy of an isomeric well. The limit of $E_{\rm exc} \rightarrow 0$ for SF is nontrivial, as the distinction between humps stops making sense when isomeric absorption is impossible. If $E_{\rm exc}$ is small, a single set of turning points $a$ and $b$ --  encompassing the entire range of $s$ for which $V(s) \geq 0$ -- suffices and the sum in Eq.~\eqref{eq:transmission_lumped} reduces to the single transition state with $E_k(J,\pi)= 0$. SF is clearly the least complex fission channel, as there is no need to model the coupling of multiple humps nor the saddle point NLD.

\emph{Microscopic fission paths} -- EDF predictions can provide an alternative to the analytical parabolic barriers and constant inertia fitted nucleus-by-nucleus to experimental data. Although the computational cost to tackle thousands of nuclei remains significant~\cite{Sanchez2026c}, mean-field EDF calculations with constraints on a collective variable have become routine. In a fission context, the most popular choice are the (dimensionless) multipole moments $\beta_{\ell m}$ of the nuclear density~\cite{Scamps21} but particle number dispersion~\cite{zdeb2025} and the neck~\cite{schunck2016} have also been explored. Taking the quadrupole moment $\beta_{20}$ as the most prominent example, one constructs a fission path by (i) setting $V(\beta_{20})$ as the EDF energy with respect to the calculated ground state and (ii) calculating $\mu(\beta_{20})$ via linear response. Nevertheless, already at this point issues emerge such as, e.g., the impact of approximations adopted to calculate the inertia~\cite{Giuliani18}, different types of discontinuities~\cite{dubray2012}, the question of symmetry restoration~\cite{marevic2020a} and the treatment of quantal corrections such as the zero-point energy~\cite{schunck2016}.

Here, we address the additional problem: EDF calculations with constraints on a set of $d$ coordinates $\mathbf{q} = (q_1, q_2, \ldots, q_d)$ produce a $d$-dimensional PES $E(\mathbf{q})$ with an associated inertia tensor $M_{ij}(\mathbf{q})$ for $i,j = 1, \ldots, d$. How can we effectively connect a multidimensional tunneling problem to the reaction codes? There is a well-studied answer to this question for SF~\cite{kapur1937,brack1972}: one constructs the one-dimensional LAP $\mathbf{q}(s)$ that minimizes the semi-classical action, taking
\begin{align}
 V(s)  & \equiv E\left[ \mathbf{q}(s) \right] - E_0\, , \\
\mu(s)&\equiv M_{\rm eff} = \sum_{i,j=1}^{d} M_{ij}(\mathbf{q}(s)) \left. \frac{dq_i}{ds} \right|_{\mathbf{q} =\mathbf{q}(s)} \left. \frac{dq_j}{ds}\right|_{\mathbf{q}=\mathbf{q}(s)} \, .
\end{align}
Here, $E_{0}$ is the energy of the local energy-minimum designated as ground state in the EDF calculations, possibly including a quantal correction, and $\mu(s)$ is often called the \emph{effective} inertia. The starting point of the path/integration is generally taken to be fixed to the ground state, while its ending point is variationally optimized within the set of outer classical turning points or among a predefined specified scission $d-1$-dimensional surface~\cite{Flynn22}. If the space of all paths has a pronounced action minimum, then the corresponding path will dominate the tunneling probability because of the exponential in Eq.~\eqref{eq:transmission_probability} and its transmission coefficient faithfully represents the $d$-dimensional tunneling problem.

Path finding is a notoriously difficult problem, even for the two-dimensional PESs that are typically explored by most EDF-based studies. Several algorithms have been proposed~\cite{baran1978,baran1981,schmid1986,mamdouh1998,Flynn22,Lemaitre2018}, but many practitioners still rely on the \emph{minimal energy path} (MEP), the path that minimizes the action under the assumption that the effective inertia is a constant. Finally, it is evident that the space of all paths is likely to have several stationary points. We label these \emph{stationary action path} (SAP); the LAP is the one with the lowest overall action among all of the SAPs.

It is natural to extend this to $E_{\rm exc} \not = 0$, i.e. to construct the fission path on the PES that maximizes the transmission coefficient $T(E_{\rm exc}, J, \pi)$ of Eq.~\eqref{eq:transmission_lumped}. Clearly, such path differs from the LAP at zero excitation energy which does not account for (i) the saddle-point NLD $\rho_s$, (ii) the variation in the classically forbidden region, i.e.~the change of the classical turning points with $E_{\rm exc}$, and (iii) any absorption in isomeric wells or coupling between humps, depending on choices made in the reaction modelling. Accounting for these effects consistently would require incorporating (a) the NLD at every point of the collective space and (b) the coupling between humps and isomers in the path finding; to the best of our knowledge, there has been no attempt at this daunting task.

Above, we silently assumed that the PES $E(\mathbf{q})$ and the tensor $M_{ij}(\mathbf{q})$ are independent of the excitation energy of the fissioning nucleus. This is crude, as both macroscopic-microscopic and EDF simulations at finite temperature predict significant decreases in barrier height and the washing out of shell effects with increasing $E_{\rm exc}$~\cite{sauer1976,schunck2015,dzhioev2025}. Although these effects impact the transmission~\cite{rahmatinejad2024}, we do not incorporate temperature here as its link with excitation energy is not without issues~\cite{schunck2015} and would anyway multiply the computational effort.

\emph{Stationary action paths with BSkG3} --
We take a first step towards consistent fission path construction at $E_{\rm exc} \not = 0$: we determine the LAP as a function of excitation energy by (i) setting $E_{\rm exc}$ to a non-zero value in Eq.~\eqref{eq:TALYS_integral}, (ii) taking \emph{just two} classical turning points close to, respectively, the EDF ground state and scission as integration limits, and (iii) clipping the integrand to zero in any isomeric region where it becomes negative. This forbids path segmentation but we do optimize the turning point close to the EDF ground state.

\begin{figure}[htbp]
\begin{center}
\includegraphics[width=\columnwidth]{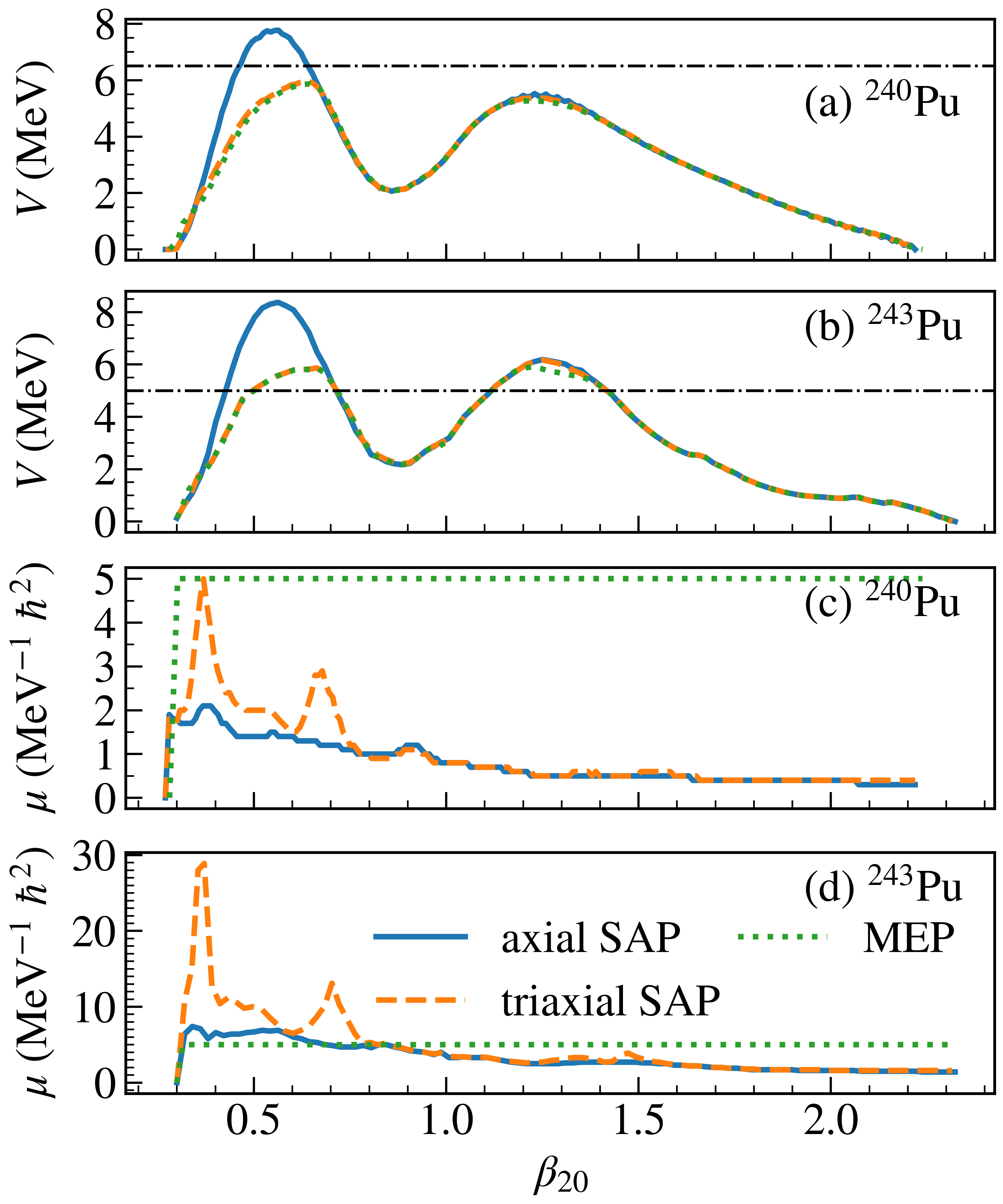}
\end{center}
\caption{\label{Pu240_Pu243_comparison}
(a)-(b) Energy profiles along the axial and triaxial SAPs and the MEP. (c)-(d) corresponding collective inertias (shown in units of $10^2$) along the different trajectories as a function of $\beta_{20}$ for $^{240}$Pu and $^{243}$Pu. We associate the MEP with the semiempirical effective inertia (see text). $S_n$ for both nuclei are indicated as black dash-dotted horizontal lines.}
\end{figure}

\begin{figure*}[htbp]
\begin{center}
\includegraphics[width=\textwidth]{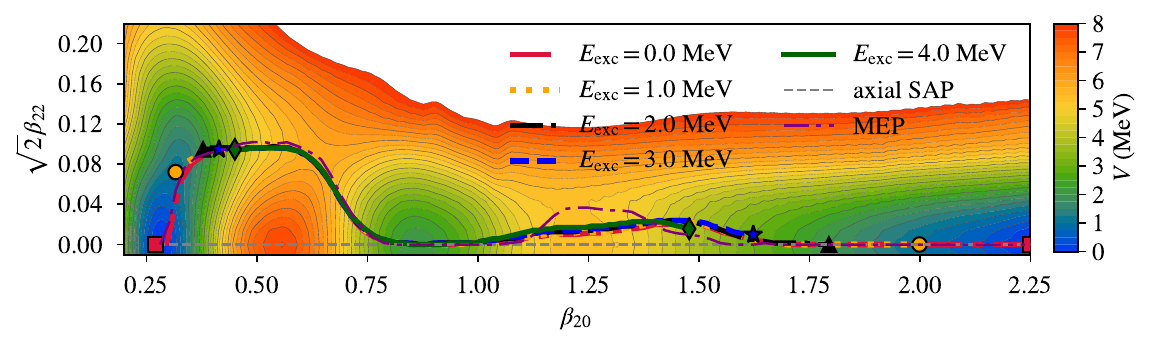}
\end{center}
\caption{\label{Pu240_diff_ein}$^{240}$Pu PES in the collective space $(\beta_{20},\beta_{22})$ obtained with BSkG3 along with triaxial SAPs at different excitation energies (in MeV). The axial SAP (dashed grey line) and the MEP (purple dash-dotted line) obtained at $E_{\rm exc}=0$ are shown for reference.}
\end{figure*}

We model fission in the two-dimensional ($\beta_{20}$, $\beta_{22}$) space, which we construct with the BSkG3 Skyrme parameterisation and the solver MOCCa~\cite{Ryssens2026} as in Ref.~\cite{Sanchez2026}, note that our approach accounts for the possibility of reflection asymmetry and the blocking effect when dealing with odd $N$ and/or $Z$. We feed these PESs and the associated ATDHFB inertia tensors $M_{ij}$ -- obtained via the cranking approximation~\cite{Baran2011,Giuliani18} -- into PyNEB~\cite{Flynn22} to obtain SAPs. At $E_{\rm exc} = 0$, this set-up describes all known SF half-lives to about three orders of magnitude \cite{Sanchez2026,Sanchez2026d}, on par with microscopic-macroscopic models.

We will focus our discussion on $^{240}$Pu and $^{243}$Pu. Our PES for $^{240}$Pu is shown in Fig.~\ref{Pu240_diff_ein}, it is qualitatively representative for nearly all actinide nuclei in our calculations. Already at $E_{\rm exc} = 0$ we find \emph{two} SAPs: the one with the lowest action proceeds through axially symmetric solutions -- the horizontal line of $\beta_{22} = 0$ in Fig.~\ref{Pu240_diff_ein} -- while another circumvents the tops of the inner ($\beta_{20} \sim 0.55$) and outer ($\beta_{20} \sim 1.2$) barriers by exploiting the triaxial $\beta_{22}$ degree of freedom. We show the mean-field energy (normalized to its minimum near $\beta_{20} \sim 0.28$) and the effective inertia along both SAPs in, respectively, the upper and lower pairs of panels of Fig.~\ref{Pu240_Pu243_comparison} for $^{240}$Pu and $^{243}$Pu. We also show the MEP and associate with it the semiempirical value of the inertia $\mu = 0.054 A^{5/3} \text{MeV}^{-1} \hbar^2 $ that is typically used in the calculation of transmission coefficients~\cite{Nilsson1969,Lemaitre2018,herman2007}. The qualitative difference in the effective inertia between $^{240}$Pu and $^{243}$Pu is due to the blocking effect of the odd neutron, which suppresses pairing correlations~\cite{rodriguez-guzman2017a}.

The coexistence mechanism of both SAPs is clearly visible: exploiting triaxiality allows the triaxial SAP to lower $V(s)$ around to the inner barrier and - to a lesser extent - the outer barrier, but movement in $\beta_{22}$ is penalized by the inertia tensor, as evidenced by the spikes in the effective inertia along the triaxial SAP. These competing effects explain why both SAPs exist $E_{\rm exc}$ for all 45 actinide nuclei for which RIPL-3 lists empirical barrier values~\cite{Capote2009}. In all cases, the actions are close: for $^{240}$Pu, approximately $43\hbar$ and $45\hbar$ for the axial and triaxial SAPs, respectively. The corresponding SF half-lives thus differ by about two orders of magnitude, but this is smaller than the typical deviation with experiment~\cite{Sanchez2026}.

\begin{figure}[htbp]
\begin{center}
\includegraphics[width=\columnwidth]{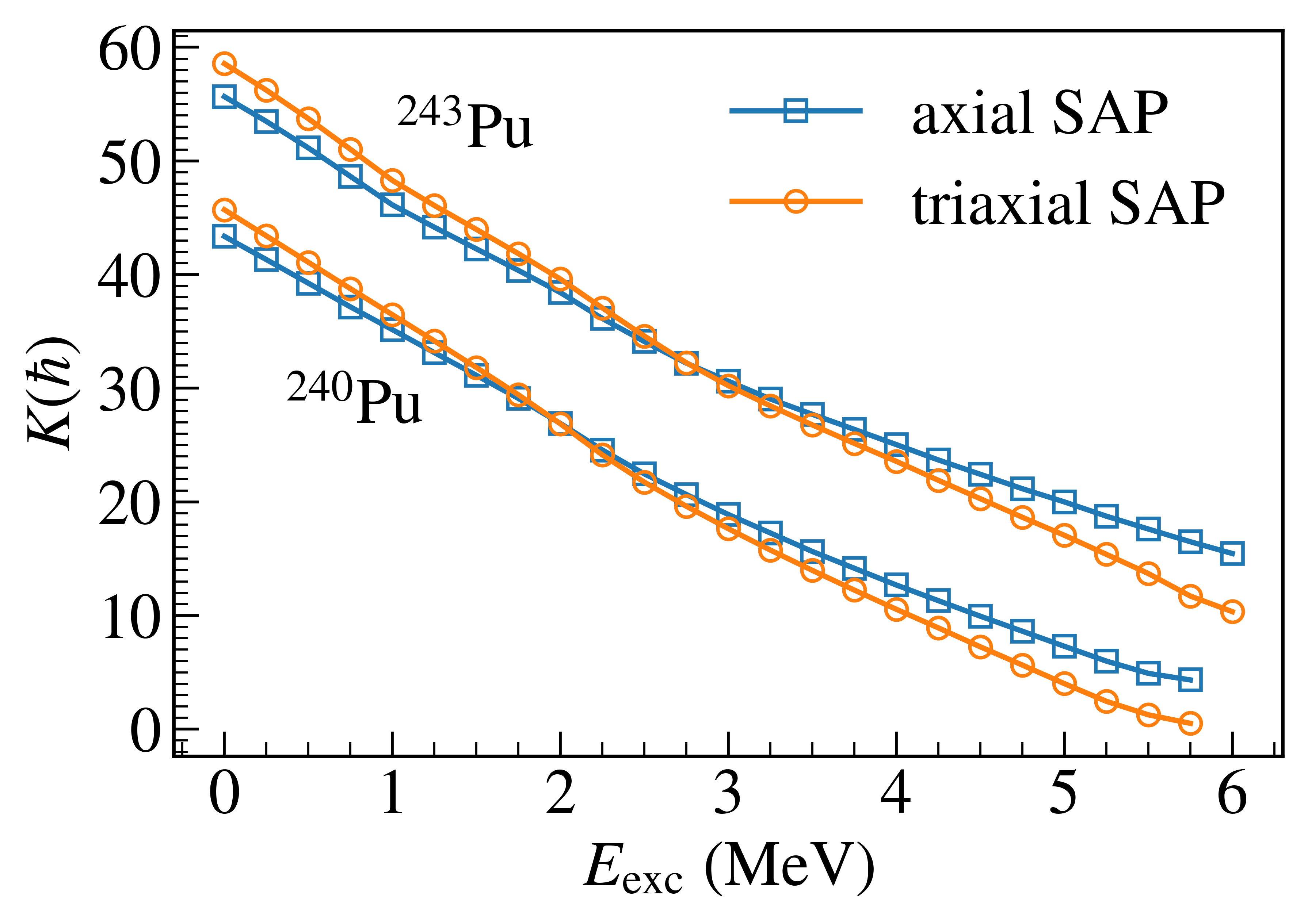}
\end{center}
\caption{\label{Pu240_Pu243_actions} Action of the SAPs of $^{240}$Pu and $^{243}$Pu as a function of the excitation energy. The curve of $^{243}\text{Pu}$ has been shifted upwards by 10$\hbar$ for ease of visualisation.}
\end{figure}

At higher excitation energy, we still find two SAPs: ``shrunk'' versions of the paths at zero excitation energy with starting/ending points at larger/smaller values of $\beta_{20}$ as the classically forbidden region shrinks with increasing $E_{\rm exc}$. We illustrate this for the triaxial SAP in Fig.~\ref{Pu240_diff_ein}.
Aside from changing turning points and small edge variations, the $(\beta_{20}, \beta_{22})$ coordinates of both SAPs are essentially independent of $E_{\rm exc}$; thus the saddle point and the associated NLD are excitation-energy independent. Evidently, the action of both SAPs decreases with increasing excitation energy as shown in Fig.~\ref{Pu240_Pu243_actions}. The rate at which this happens is different for both paths: the axial SAP has the lowest action at low $E_{\rm exc}$ but this ordering inverts at a moderate excitation energies, approximately 1.75 and 2.75~MeV for $^{240}$Pu and $^{243}$Pu. We label the transition energy $E_{\rm crit}$. That the action of both SAPs behaves differently with excitation energy is easily understood from Fig.~\ref{Pu240_Pu243_comparison}: the segments of the triaxial SAP where the effective inertia is high (where $\beta_{22}$ varies) are progressively excluded from the classically forbidden region as $E_{\rm exc}$ increases.

\begin{figure}[htbp]
\begin{center}
\includegraphics[width=\columnwidth]{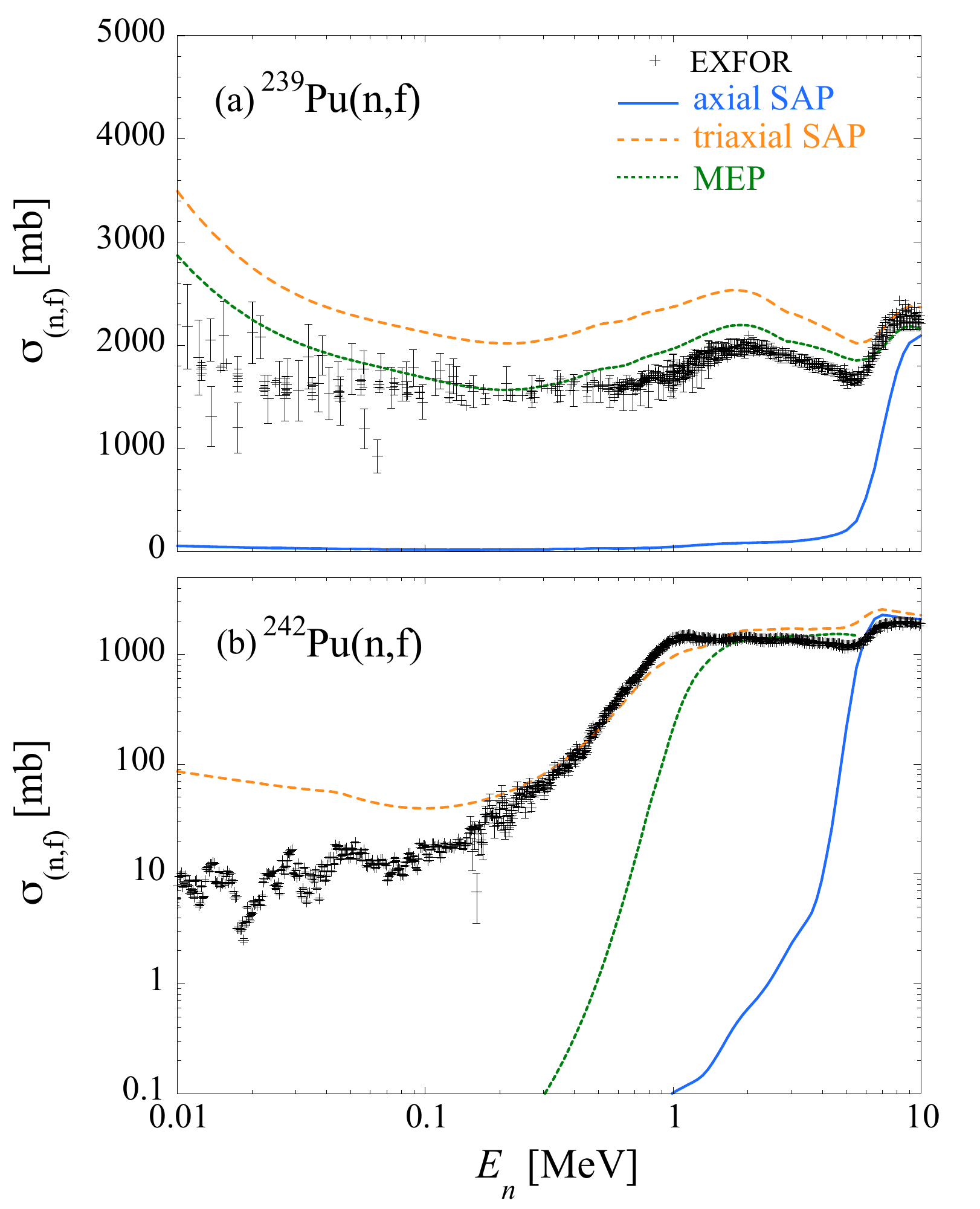}
\end{center}
\caption{\label{cross_sections} Comparison of the calculated (a) $^{239}$Pu$(n,f)$ and (b) $^{242}$Pu$(n,f)$ cross sections as a function of the incident neutron energy, obtained with TALYS using the axial and triaxial SAPs and the MEP. The experimental data is from EXFOR \cite{Otuka2014}.}
\end{figure}

\emph{Neutron-induced fission cross sections} We feed the MEP and both SAPs, determined at $E_{\rm exc} = 0$, and the corresponding saddle-point NLDs, determined entirely consistently based on the combinatorial method~\cite{goriely2026}, into TALYS. The resulting nF cross sections -- without any additional renormalization --  are shown in Fig.~\ref{cross_sections}. In both cases, it is \emph{not} the LAP at zero $E_{\rm exc}$ which offers the best description of the data; instead, it is the paths that exploit the triaxial degree of freedom. Much of the variation in cross section is due to the difference in primary fission barrier height: for both isotopes, triaxiality lowers the inner barrier by about 2 MeV. For $^{239}$Pu$(n,f)$, the compound nucleus $^{240}$Pu is formed at $Sn \approx 6.5$ MeV: lower than the primary barrier along the axial SAP but above that of the triaxial paths. TALYS uses Eq.~\eqref{eq:P_hillwheeler} above the barrier, hence the remaining modest difference between MEP and triaxial SAP is essentially due to the difference in effective inertia at the saddle point. For $^{242}$Pu$(n,f)$, the compound nucleus $^{243}$Pu is formed at an excitation energy of $S_n \approx 5.0$~MeV~\cite{Wang2021}, below the barrier for all paths. Here, the remaining variation between MEP and the triaxial LAP is determined mostly by the difference in effective inertia along the path. However, BSkG3 was constructed with an adjustment of the barrier heights along the MEPs to the empirical barriers of RIPL-3~\cite{grams23}; we thus do not claim that the experimental data on cross sections favors triaxial paths. Our conclusion is more modest: multiple SAPs can coexist and lead to different nF cross sections even if the corresponding SF half-lives match experiment equally well.

\begin{figure}[htbp]
\begin{center}
\includegraphics[width=\columnwidth]{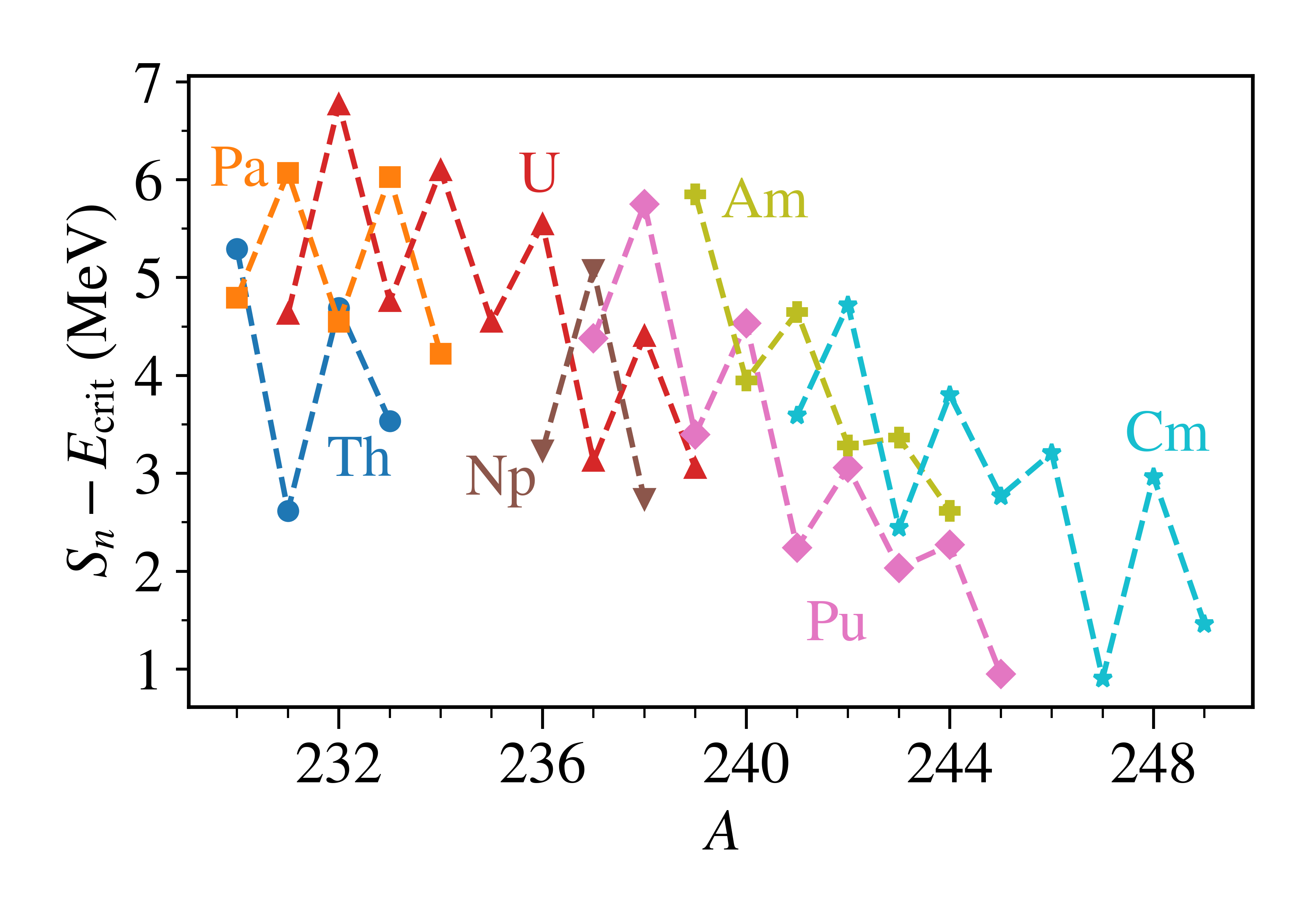}
\end{center}
\caption{\label{sn-e_crit_ripl3} Difference between the experimental neutron separation energy of the compound nucleus, $S_n$, and the energy at which the LAP goes through triaxial barriers as obtained with BSkG3 for the 45 actinides with empirical RIPL-3 barriers.}
\end{figure}

\begin{figure}[htbp]
\begin{center}
\includegraphics[width=\columnwidth]{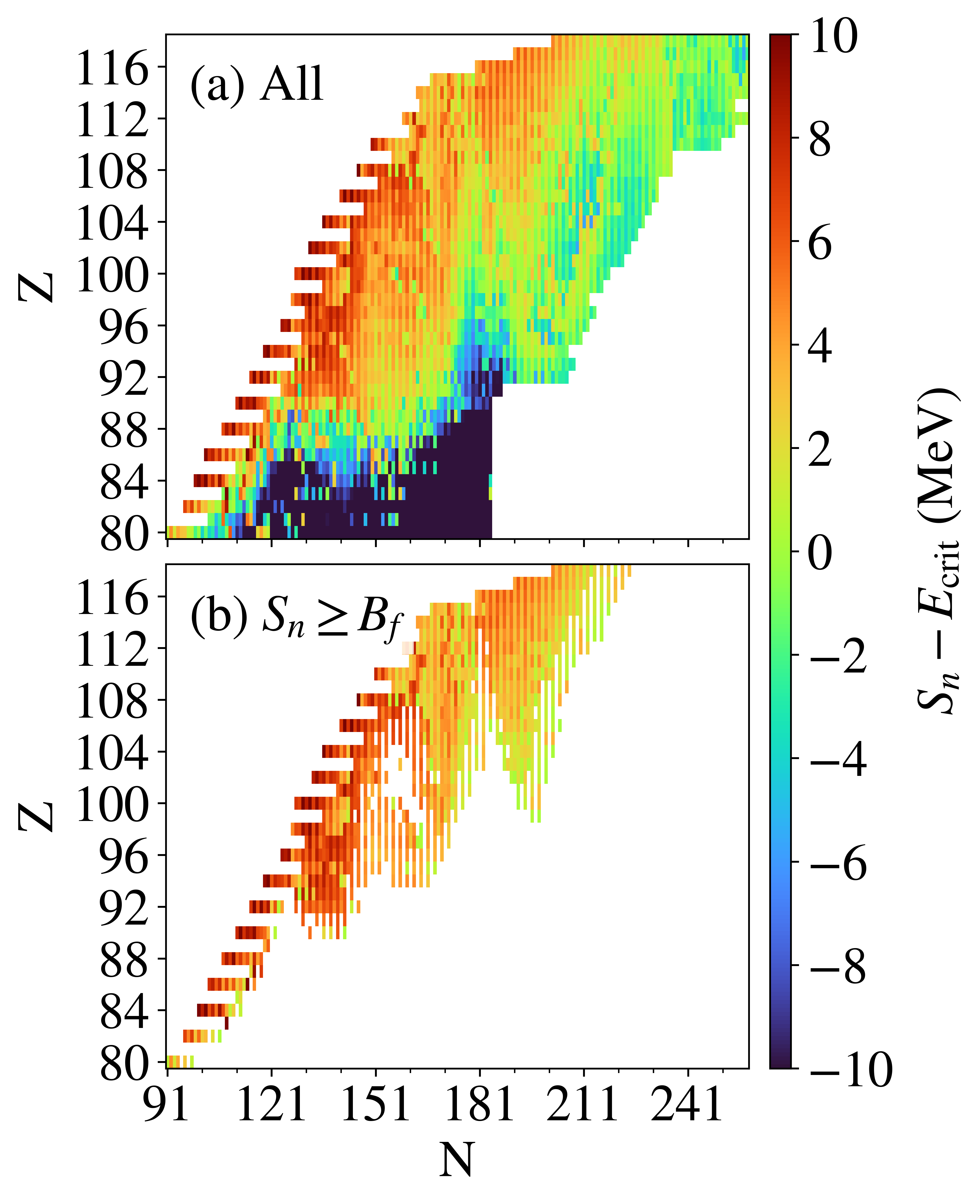}
\end{center}
\caption{\label{NZ_plane} Difference $S_n-E_{\rm crit}$ for (a) all $Z \geq 80$ nuclei within the BSkG3 driplines and (b) the subset for which $S_n$ exceeds the primary fission barrier along the triaxial SAP.}
\end{figure}

\emph{Global analysis} --  We extended this analysis to the 45 RIPL-3 actinides~\cite{Capote2009}: in every case, the triaxial SAP is not the LAP at zero excitation energy. Fig.~\ref{sn-e_crit_ripl3} shows $S_n-E_{\rm crit}$ in: in all cases, the triaxial SAP is the LAP at the excitation energy of the compound nucleus in nF. On a larger scale, panel (a) of Fig.~\ref{NZ_plane} shows this difference for all $Z \geq 80$ nuclei within the BSkG3 drip lines; panel (b) shows the subset of nuclei cases for which nF is expected to occur even in the thermal-neutron energy regime, i.e. fissile nuclei for which $S_n$ is larger than the primary fission barrier. The global picture is clear: at the excitation energies relevant to nF, the triaxial SAP is favoured for essentially all nuclei so $r$-process fission data should account for this. Only for pre-actinides and some of the most neutron-rich actinides with $Z \leq 94$ is $E_{\rm crit} \ge S_n$; however, (i) BSkG3 predicts many of these to have triaxial ground states such axial vs.~triaxial SAP comparison is not straightforward~\cite{grams23,Sanchez2026b} and that (ii) many $Z \leq 90$ nuclei have high barriers, such that nF differences between SAPs need not be as dramatic as for $^{240,243}$Pu.

\textit{Conclusions and outlook} -- We have analyzed the mismatch between the microscopic fission studies and the assumptions of reaction codes for fission cross section calculations. With the BSkG3 PESs of two Pu isotopes, that exploits consistently both the triaxial and octupole degrees of freedom,
we demonstrated that multiple one-dimensional stationary action paths can coexist and that their action ordering can change as a function of excitation energy. Since different fission channels operate at different excitation energy, the appropriate path for a reaction code need not be the same for spontaneous, nF or $\beta$-delayed fission. In particular, such paths might lead to SF half-lives that match known data equally well but yield dramatically different nF cross sections. For BSkG3, our global analysis shows that nF cross sections should systematically use the triaxial SAP~\cite{Sanchez2026b}. This conclusion and other details of our analysis likely depend on modelling choices such as, e.g., our choice of parameterisation, collective coordinates and the treatment of collective inertia~\cite{Sadhukhan13}. Nevertheless, our results show one cannot blindly assume that the least action path is appropriate for the calculation of fission transmission coefficients.

Achieving a perfect match between reaction codes and structure theory will be difficult. On the reaction side, transmission coefficients could be determined in multiple dimensions via an absorbing potential~\cite{scamps2015a} or multidimensional transition state theory~\cite{wigner1938}; doing so would mean abandoning the JWKB approximation but would have the advantage of accommodating multimodal fission~\cite{dubray2008,Goriely2013,regnier2019,lay2024a}. A more pragmatic solution would switch between multiple fission paths, possibly determined by the consistent maximisation of $T(E_{\rm exc})$, depending on excitation energy. Our findings also affect the future construction of EDF models: because empirical barriers are determined by fitting to nF data, fit protocols should not blindly match them to highest point of the calculated LAPs if the goal is to predict cross sections. The extension of our results to $\beta$-delayed fission probabilities is under way.

\appendix

\section*{Acknowledgments}

This work was supported by the Fonds de la Recherche Scientifique (F.R.S.-FNRS) and the Fonds Wetenschappelijk Onderzoek--Vlaanderen (FWO) under the EOS Projects No.~O000422 and O022818F, as well as by the F.R.S.-FNRS under the MIS Project No.~40028446. This research benefited from computational resources made available on the Tier-1 supercomputer Lucia of the Fédération Wallonie--Bruxelles, infrastructure funded by the Walloon Region under Grant Agreement No.~1117545. Additional computational resources were provided by the Consortium des Équipements de Calcul Intensif (CÉCI), funded by F.R.S.-FNRS under Grant No.~2.5020.11 and by the Walloon Region. We acknowledge the EuroHPC Joint Undertaking for awarding access to the MareNostrum5 supercomputer at the Barcelona Supercomputing Center (BSC), Spain, under projects EHPC-DEV-2025D04-097 and EHPC-DEV-2026D01-043. We further acknowledge EPICURE, a EuroHPC Joint Undertaking initiative, for supporting our work on MareNostrum5 in the context of these development projects.

\bibliography{apssamp,fission_LAPs}

\end{document}